\documentclass[journal]{IEEEtran}

\usepackage[utf8]{inputenc}
\usepackage[T1]{fontenc}
\usepackage{amsmath,amssymb,amsfonts}
\usepackage{booktabs}
\usepackage{graphicx}
\usepackage{multirow}
\usepackage{xcolor}
\usepackage{xspace}
\usepackage{algorithm}
\usepackage{algpseudocode}
\usepackage[hidelinks]{hyperref}
\usepackage{cleveref}
\crefname{figure}{Fig.}{Figs.}
\Crefname{figure}{Fig.}{Figs.}

\newcommand{\auc}{\mathrm{AUC}}
\newcommand{\pauc}{\mathrm{pAUC}}
\newcommand{\source}{\mathrm{s}}
\newcommand{\target}{\mathrm{t}}
\newcommand{\method}{\textsc{DACo}\xspace}

\begin{document}

\title{Training-Free Model Selection and Domain-Aware Score Calibration for
First-Shot Anomalous Sound Detection}

\author{Grach~Mkrtchian%
\thanks{G.~Mkrtchian is an independent researcher
(e-mail: g.mkrtchyan.m@gmail.com).}}

\maketitle

\begin{abstract}
First-shot anomalous sound detection in DCASE Challenge Task~2 must flag
anomalies of unseen machine types with a single threshold, without knowing
whether a test clip comes from the data-rich \emph{source} domain (990
normal training clips) or the data-scarce \emph{target} domain (10). Two
organizer-reported problems remain open: source- and target-domain AUC are
negatively correlated across systems, and development-set performance does
not predict evaluation-set performance. We address both with a
training-free post-hoc layer over frozen audio embeddings:
(i)~per-domain quantile calibration shrunk toward a pooled map by a prior
strength $m$, tracing a controllable source/target balance frontier, and
(ii)~a \emph{label-free cross-validated domain-balance criterion} that
ranks candidate configurations from training normals only, paired with a
coarse development-labeled viability veto. On DCASE 2025, the criterion
rank-predicts the official evaluation score across a 45-configuration grid
(Spearman $\rho_s{=}{+}0.91$; family-block bootstrap 95\% CI
$[+0.83,+0.95]$) while development score is uninformative ($+0.06$, CI
$[-0.39,+0.31]$). Criterion-based selection raises evaluation $\Omega$
from $55.83$ to $59.34$ (jackknife CI $[2.2,4.8]$) and, on an extended
grid, to $61.05$---retrospectively fourth of 35 teams. Replication
on DCASE 2023 and 2024 bounds the claim: development score is
uninformative in all three years and well-balanced degenerate
configurations recur (vetoed in every case), but under family-clustered
uncertainty the criterion's predictive evidence survives only in 2025;
a fixed full-equalization default matches or beats criterion selection in
both replication years, and the selection gain over development-based
selection is significant only in 2025 ($+5.2$, jackknife CI $[1.3,9.2]$).
A DCASE 2026 forward test is frozen before the per-clip evaluation ground
truth becomes available; all headline numbers are reproduced by the
official evaluator.
\end{abstract}


\section{Introduction}\label{sec:intro}
\IEEEPARstart{M}{achine} condition monitoring by sound must operate on
machine types the developer has never recorded, be deployed with almost no
data from the operating environment, and run at a fixed sensitivity chosen
before any anomaly has been observed. DCASE Challenge Task~2 crystallizes
this setting as \emph{first-shot unsupervised ASD under domain
generalization}~\cite{dcase2025task2,harada2023firstshot}: for each machine
type, training data comprise 990 normal clips from a \emph{source} domain and
only 10 from a \emph{target} domain; at test time the domain label is
withheld, so anomalies from both domains must be separated from normal sounds
with a \emph{single} threshold, and the evaluation machine types are disjoint
from the development (henceforth ``dev'') ones, forbidding per-machine
tuning.

The 2025 organizers document two failure modes that remain
open~\cite{dcase2025task2}. First, \emph{source/target imbalance}: across the
top twenty teams, source- and target-domain AUC are negatively correlated,
and only four teams beat the official baselines in both domains
simultaneously. Second, \emph{dev$\to$evaluation non-transfer}: ``achieving
high AUC values in the development dataset does not indicate high AUC in the
evaluation dataset''---model selection on the dev set is unreliable.

This article argues that both failure modes are, at their core, calibration
and model-selection problems, and that they can be attacked \emph{post hoc},
on top of any frozen embedding extractor and any training-free anomaly-score
backend, at negligible computational cost. The raw anomaly scores of source-
and target-domain normal clips live on different scales, so no single
threshold can be well placed for both; and the dev set is the wrong yardstick
for choosing how strongly to correct this mismatch.

Our proposal, \method (from \emph{Domain-Aware Calibration}), has
two parts whose importance our experiments rank in this order:

\emph{A label-free selection criterion} (\cref{sec:stage3}). For any
candidate configuration, repeatedly hold out half of each domain's training
normals, refit the score pipeline, and measure the Kolmogorov--Smirnov (KS)
distance between the calibrated held-out source and target score
distributions. The criterion measures how domain-balanced the configuration's
operating point is, uses no anomalies and no test data, and is therefore
available for unseen machines at deployment. It is family-agnostic: it
selects among raw, normalized, and calibrated scoring pipelines alike.

\emph{A per-domain calibration layer with a controllable strength}
(\cref{sec:method}). Test clips receive soft latent-domain weights from
embedding-space proximity to the source and target training banks; the base
score is mapped through per-domain quantile maps built from leave-one-out
training scores; each map is shrunk toward the pooled map by a prior strength
$m$, tracing a smooth frontier between the raw ranking ($m{\to}\infty$) and
full per-domain equalization ($m{=}0$). The frontier turns ``how strongly to
calibrate'' into a selectable quantity---the object the criterion selects.

Our main findings on DCASE 2025 Task~2 are:
\begin{enumerate}
\item \emph{Dev and evaluation scores disagree structurally}
(\cref{sec:results-transfer}): across a 45-configuration grid, dev $\Omega$
(the official harmonic-mean metric) increases with $m$ (Spearman
$\rho_s{=}{+}0.45$) while evaluation $\Omega$ decreases ($-0.96$); their
overall rank correlation is $+0.06$. This is a sharper, mechanism-level
restatement of the organizers' transfer observation.
\item \emph{The label-free criterion transfers}
(\cref{sec:results-transfer}): computed on dev machines, it rank-predicts
evaluation $\Omega$ at $\rho_s{=}{+}0.91$ (family-block bootstrap 95\% CI
$[+0.83,+0.95]$; machine-level bootstrap $[+0.24,+0.97]$). Guarded selection
by the criterion instead of dev $\Omega$ raises evaluation $\Omega$ from
$55.83$ (below the official Selective-Mahalanobis baseline; rank 23 when inserted
into the 35-team leaderboard) to $61.05$ (rank 4; extended grid,
\cref{sec:results-ldn})---with the identical training-free system pool.
\item \emph{The calibration frontier controls the aggregate source/target
trade-off} (\cref{sec:results-frontier}): the signed mean gap moves from
$+13.8$ points (raw) to $-1.2$ ($m{=}0$, a slight target overshoot). We
disclose that per-machine balance is a different story (mean absolute gap
moves only from $15.1$ to $14.0$), and analyze the failure cases.
\item \emph{The pattern is robust where it must be, and we map where it is
not} (\cref{sec:results-ablations,sec:results-ldn,sec:results-years}): on
three frozen backbones the criterion predicts evaluation rank at
$\rho_s{=}{+}0.67$ to $+0.81$ while dev $\Omega$ swings between $-0.03$ and
$+0.84$. Local-density score
normalization~\cite{wilkinghoff2025localdensity} is itself strong on
evaluation ($61.04$) but \emph{rejected} by dev-based selection ($56.5$ dev
vs.\ $58.7$ raw); the criterion selects into that family. Replicating the
full pipeline on DCASE 2023 and 2024 turns the single dev$\to$eval pair
into three: dev $\Omega$ is uninformative in every year, well-balanced
degenerate configurations recur in every year (the unguarded criterion
selects one in three of five year--backbone settings; the veto catches all
of them), and the criterion's raw rank correlations are $+0.54$, $-0.10$,
and $+0.81$ (like-for-like 51-configuration grids)---but only the 2025
evidence survives family-clustered uncertainty analysis
(\cref{sec:results-years}), so we claim criterion transfer for one year
of three and treat 2023 as suggestive at most.
\item \emph{A genuinely prospective test is pre-registered}
(\cref{sec:results-years}): the DCASE 2026 evaluation data do not yet
exist, so we froze the criterion-selected configuration on the released
2026 development set---selection, veto, channel policy, and artifact hashes
committed to the public repository---before any outcome can be known.
\item \emph{Everything is training-free and cheap}
(\cref{sec:results-efficiency}): embedding extraction runs once
($1.9$--$9.4$\,ms/clip); a full configuration cycle on cached embeddings,
criterion included, takes ${\sim}0.15$\,s per machine.
\end{enumerate}

We also contribute a negative-result analysis: pure per-domain calibration
($m{=}0$) \emph{harms} the dev set on specific machine geometries, the
criterion does not predict \emph{dev} performance, a purely label-free
criterion can be gamed by degenerate score maps---which motivates the
coarse dev-side viability veto we pair it with---and in two of three
challenge years a fixed full-equalization default matches or beats
criterion-based selection, concentrating the criterion's demonstrated
added value in a single year. That the same
calibration knob helps evaluation machines and hurts dev machines \emph{is}
the organizers' transfer phenomenon; we show it can be navigated with
deployment-legal signals, and we bound where the navigation helps.

\section{Related Work}\label{sec:related}

\subsection{Systems for domain-generalized ASD}
The dominant DCASE Task~2 recipe learns discriminative embeddings with
angular-margin auxiliary classification: sub-cluster
AdaCos~\cite{wilkinghoff2021subcluster}, self-supervised
FeatEx~\cite{wilkinghoff2023featex}, and subspace-projection
AdaProj~\cite{wilkinghoff2024adaproj}; see the recent review of
DCASE-related domain-shift work~\cite{wilkinghoff2025review} for a taxonomy.
Frozen audio foundation models (BEATs~\cite{chen2022beats},
EAT~\cite{chen2024eat}, PANNs~\cite{kong2020panns}) with training-free
backends are competitive: GenRep~\cite{genrep2025}---frozen BEATs + kNN with
source/target memory banks, target-bank augmentation, and domain score
normalization---placed second among teams in DCASE 2025 Task~2, and a recent
systematic study finds the scoring backend matters more than
pooling~\cite{zhou2026backends}. Our layer operates on top of exactly this
system class. The official Selective-Mahalanobis baseline likewise keeps
per-domain statistics~\cite{dcase2025task2}; per-domain \emph{modeling} is
standard---what is new here is pinning both domains' normal scores to a
common quantile scale and treating the strength of that correction as a
selectable quantity.

\subsection{Score normalization}
The closest prior line is score normalization for domain generalization:
ratio- and difference-based rescaling by local density of the reference
set~\cite{wilkinghoff2025icassp,wilkinghoff2025localdensity}, with a
cluster-exit extension~\cite{wilkinghoff2026clusterexit}. Normalization makes
scores \emph{comparable} across acoustic regions; per-domain quantile
calibration additionally anchors them to an \emph{operating point}
(approximate per-domain false-positive rate, empirically validated in
\cref{sec:results-frontier}). More importantly, that line does not treat
dev$\to$eval transfer or model selection, which our experiments show is
where the larger gain lies; indeed, dev-based selection would reject
local-density normalization itself (\cref{sec:results-ldn}).

\subsection{Conformal and group-conditional calibration}
Conformal anomaly detection yields distribution-free $p$-values from
calibration scores~\cite{vovk2005algorithmic,laxhammar2015,
angelopoulos2021conformal}. Our Stage-2 maps are Mondrian (group-conditional)
in spirit~\cite{vovk2005algorithmic}, with the domain as the taxonomy;
borrowing strength between small groups and a pooled calibration set connects
to class-conditional conformal prediction with many
classes~\cite{ding2023classconditional} and conditional-guarantee
work~\cite{gibbs2023conditional}. Recent applications to industrial and
time-series monitoring include adaptive conformal detection on frozen
time-series foundation models~\cite{ibm2026adaptiveconformal}, weighted
conformal $p$-values in low-data regimes~\cite{weightedconformal2026}---whose
granularity analysis directly bears on our 10-clip target domain---and
conformal false-alarm control for cyber-physical
systems~\cite{ress2026conformal}. Set-valued prediction pursues per-domain
criteria for supervised domain generalization~\cite{setvalued2025}.
Weighting calibration data by proximity to the test point is itself an
established idea---localized conformal
prediction~\cite{guan2023localized} and randomly-localized weighting
with robust guarantees~\cite{hore2023localweights}---so our soft
latent-group membership should be read as a two-group special case of
that principle rather than a new mechanism; likewise, the shrinkage form
of \cref{eq:shrink} is the classical $m$-estimate / partial-pooling
estimator~\cite{cestnik1990,gelman2007arm} applied to quantile maps, and
differs from the weighted-conformal construction
of~\cite{weightedconformal2026} in mixing \emph{maps} rather than
reweighting calibration points. We are not aware of prior work applying
per-domain FPR anchoring of this kind to ASD; the method-level
contribution is treating the pooling strength $m$ as a selection
axis---the frontier---rather than any single ingredient.

\subsection{Label-free model selection for anomaly detection}
Selecting anomaly detectors without labels is an open problem with its own
literature: meta-learning from prior labeled collections
(MetaOD~\cite{metaod2021}) and surrogate-signal ranking for time-series AD
(\cite{goswami2023uoms}). In ASD specifically, \cite{zhou2026backends}
report a \emph{failed} pseudo-validation selection experiment, underscoring
the gap. A parallel lineage exists for unsupervised \emph{domain
adaptation}, where models must likewise be chosen without target labels:
importance-weighted cross-validation~\cite{sugiyama2007iwcv}, Deep
Embedded Validation~\cite{you2019dev}, and soft neighborhood
density~\cite{saito2021snd} construct label-free surrogates of target
risk for classifiers. Our criterion differs in target and signal: it
selects for \emph{domain balance of the operating point} of an anomaly
detector in the first-shot single-threshold setting, using only the
machine's own normal training audio, and is validated by dev$\to$eval
transfer against the official protocol. To our knowledge no prior work
proposes a label-free selection signal for this specific
setting---operating-point balance under first-shot domain
generalization---though the surrogate-validation idea itself is
well established.

\section{Problem Setting}\label{sec:setting}

\subsection{Data and protocol}
We use DCASE 2025 Challenge Task~2~\cite{dcase2025task2}. The dev set
contains seven machine types (\texttt{fan}, \texttt{gearbox},
\texttt{bearing}, \texttt{slider}, \texttt{valve}, \texttt{ToyCar},
\texttt{ToyTrain}); the evaluation set contains eight novel types
(\texttt{AutoTrash}, \texttt{BandSealer}, \texttt{CoffeeGrinder},
\texttt{HomeCamera}, \texttt{Polisher}, \texttt{ScrewFeeder}, \texttt{ToyPet},
\texttt{ToyRCCar}). Per machine type: 990 source-domain and 10 target-domain
normal training clips (domain labels available for training data), and 200
test clips (100 normal, 100 anomalous; both domains) whose domain and
condition are withheld at inference. All audio is single-channel 16\,kHz,
built from ToyADMOS2~\cite{harada2021toyadmos2} and
MIMII~DG~\cite{dohi2022mimiidg}. Post-challenge ground truth for the
evaluation set is published in the official evaluator repository, enabling
offline scoring; we use it strictly for metric computation, never for
selection.

\subsection{Metrics}
Following the official protocol, $\auc_{j,d}$ is computed per machine $j$ and
domain $d\in\{\source,\target\}$ using the normal test clips of domain $d$
against \emph{all} anomalous clips of both domains; $\pauc_j$ is the
standardized partial AUC over the false-positive range $[0,0.1]$ over all
test clips; the official score is the harmonic mean over machines and
quantities,
\begin{equation}
\Omega=\mathcal{H}\big(\{\auc_{j,\source},\auc_{j,\target},\pauc_j\}_{j\in
\mathcal{M}}\big),
\label{eq:omega}
\end{equation}
reported $\times100$ throughout. The harmonic mean punishes imbalance, which
is why the source/target trade-off is damaging. Our implementation of
\cref{eq:omega} matches the official evaluator: a 2000-trial randomized
differential test against the evaluator's per-domain filtering logic agrees
to machine precision (script in the released code), and all headline
evaluation-set numbers were additionally reproduced end-to-end by running
the official evaluator on exported per-clip scores
(\cref{sec:results-transfer}).

\section{Method}\label{sec:method}

\method operates after a frozen embedding extractor
$\phi:\mathcal{X}\to\mathbb{R}^D$ and a base anomaly scorer $A$; neither is
modified. Let $\mathcal{B}=\{\phi(x_i)\}_{i=1}^{N}$ be a machine's normal
training embeddings ($N{=}1000$), partitioned by the known training-domain
labels into $\mathcal{B}_\source$ ($n_\source{=}990$) and
$\mathcal{B}_\target$ ($n_\target{=}10$).

\subsection{Preliminaries: base scores and calibration scores}
We use the $k$-nearest-neighbor cosine distance as the base score,
$A(x)=\tfrac1k\sum_{z\in\mathrm{NN}_k(x;\mathcal{B})}d_c(\phi(x),z)$, with
$d_c$ the cosine distance on $\ell_2$-normalized embeddings. Calibration
scores for training clips are computed \emph{leave-one-out} (LOO): the score
of $x_i\in\mathcal{B}$ is its mean distance to its $k$ nearest neighbors in
$\mathcal{B}\setminus\{x_i\}$, mirroring test-time scoring while excluding
the trivial self-match.

\subsection{Stage 1: latent-domain assignment}\label{sec:stage1}
The test-clip domain is unknown. Let
$d_\source(x)=\min_{z\in\mathcal{B}_\source}d_c(\phi(x),z)$ and analogously
$d_\target(x)$. Soft weights are a softmax over $-d_g(x)/T$,
$g\in\{\source,\target\}$, with temperature $T$ set to the median of the
machine's pooled LOO calibration scores (the typical neighbor-distance
scale); a hard variant uses $w_\target=\mathbf{1}[d_\target<d_\source]$.
Stage~1 uses only the machine's own training data, respecting the first-shot
constraint.

\subsection{Stage 2: per-domain quantile calibration with shrinkage}
\label{sec:stage2}
For each domain $g$, let $\hat{c}_g$ be the \emph{interpolated} empirical CDF
of the LOO calibration scores of $\mathcal{B}_g$: order statistics at
plotting positions $i/(n_g{+}1)$, linearly interpolated, with a monotone
rational right tail $q_{\max}+(1-q_{\max})\,\xi/(\xi+\lambda)$, where
$q_{\max}=n_g/(n_g{+}1)$, $\xi$ is the exceedance over the largest
calibration score $s_{\max}$, and
$\lambda=\max(s_{\max}-\mathrm{med},10^{-9})$ with $\mathrm{med}$ the
median calibration score---so extreme scores retain their raw ordering
rather than saturating; below the smallest calibration score the map
interpolates linearly toward zero at score $0$. Interpolation avoids the $1/(n_\target{+}1)$-granularity ties a
step CDF would produce with
$n_\target{=}10$~\cite{weightedconformal2026}.

Because ten points cannot support extreme quantiles, each domain map is
shrunk toward the pooled map $\hat{c}_{\mathrm{pool}}$ (all $N$ LOO scores)
with prior strength $m\ge 0$:
\begin{equation}
c_g(a)=\frac{n_g\,\hat{c}_g(a)+m\,\hat{c}_{\mathrm{pool}}(a)}{n_g+m}.
\label{eq:shrink}
\end{equation}
The calibrated score is $\tilde A(x)=\sum_g w_g(x)\,c_g(A(x))$.

\emph{What is and is not guaranteed.} At $m{=}0$ each domain's normal scores
map approximately to $\mathrm{U}(0,1)$, so one global threshold $\tau$
corresponds to a similar normal-exceedance rate $1-\tau$ in every latent
domain. We deliberately call this \emph{approximate} FPR equalization rather
than a conformal guarantee: the LOO construction is jackknife-style rather
than split-conformal (calibration scores are mutually dependent), linear
interpolation forgoes the exactness of step-function conformal $p$-values,
soft assignment mixes the two maps, any Stage-1 error breaks per-domain
exchangeability, and for $m{>}0$ no guarantee is intended. The map also has
data support only down to exceedance $1/(n_g{+}1)$; with $n_\target{=}10$,
all of the $\pauc$ integration range except the sliver $[1/11,\,1/10]$ lies in
the extrapolated tail. We therefore validate the property empirically: at
matched label-free thresholds, calibration roughly halves the cross-domain
FPR imbalance relative to raw scoring
(\cref{sec:results-frontier}). As $m\to\infty$ both maps coincide with
$\hat{c}_{\mathrm{pool}}$, a single monotone transform preserving the raw
ranking; $m$ thus \emph{parametrizes the source/target balance frontier}
(\cref{fig:frontier}), turning calibration strength into a model-selection
problem.

\subsection{Stage 3: the selection criterion}\label{sec:stage3}
Dev-set $\Omega$ is the standard selection signal, but it does not transfer
(\cref{sec:results-transfer}). We select among configurations (here: $m$,
$k$, assignment, calibration variant, score family) with a criterion
computable from \emph{training normals only}. For each of $S$ random splits
(indexed $r$),
hold out half of each domain's training normals, refit the pipeline
(Preliminaries--Stage~2) on the remainder, push the held-out normals through
it, and compute
\begin{equation}
C=\tfrac1S\sum_{r=1}^{S}
\mathrm{KS}\big(\tilde A(\mathcal{D}^r_\source),\,
               \tilde A(\mathcal{D}^r_\target)\big),
\label{eq:criterion}
\end{equation}
where $\mathcal{D}^r_g$ is the held-out half of domain $g$ in split $r$.
Equation~\eqref{eq:criterion} is computed per machine; for
fixed-configuration selection we aggregate as the mean over the dev
machines, $C(\gamma)=|\mathcal{M}_{\mathrm{dev}}|^{-1}\sum_{j}
C_j(\gamma)$, while per-machine selection uses the target machine's own
$C_j$. Small $C$ means the two domains' normal score distributions nearly coincide,
so one global threshold produces similar false-positive rates in both---the
domain-balanced operating point that the harmonic-mean metric rewards. The
criterion is defined for every configuration including the uncalibrated
baseline, and uses no anomalies, no test clips, and no cross-machine
statistics. We use $S{=}10$ with a fixed seed pairing splits across
configurations; the selected configuration is unchanged for
$S\in\{5,10,20\}$. Note the finite-sample floor: with only 5 held-out target
clips per split, the two-sample KS statistic under identical distributions
averages ${\approx}0.36$ per machine (${\pm}0.012$ for a 7-machine mean), so
observed values ($0.49$--$0.72$) must be read relative to that floor, and
differences of ${\lesssim}0.01$ between configurations are noise---the
criterion is a coarse region selector, not a fine ranker
(\cref{sec:results-transfer}). One further subtlety: during criterion
CV the pipeline is refit on half-size banks ($n'_g=n_g/2$, so
$n'_\target{=}5$), which makes the effective pooled weight in
\cref{eq:shrink} $m/(n'_g{+}m)$ at selection time versus $m/(n_g{+}m)$
at deployment---slightly stronger shrinkage during selection. At
$m{=}0$, where all headline picks lie, the two coincide; for $m{>}0$
the mismatch could be removed by rescaling $m'=m\,n'_g/n_g$, which we
note as a limitation rather than adopt.

\subsection{Viability veto}\label{sec:veto}
Balance is necessary but not sufficient: a degenerate score map can be
perfectly balanced and detect nothing, and a label-free criterion cannot see
detection power (a concrete example arises in \cref{sec:results-ldn}). Dev
$\Omega$, although unreliable for fine ranking, is reliable at flagging
broken configurations: the degenerate maps in our grids fall $8$--$16$
points below the dev-best. The full selection rule
(\cref{alg:daco}) is therefore: \emph{exclude the bottom decile of
configurations by dev $\Omega$, then minimize the criterion}. A threshold
form of the veto (within $\delta$ of the dev-best) gives identical
selections for $\delta\in[3,8]$ on two of three backbones; we disclose the
exceptions, and the veto's origin, in \cref{sec:results-ablations}. The
veto uses labeled dev data only---legal before any evaluation data exist.
The complete rule is thus label-free \emph{on the deployment machines};
dev labels enter only as a coarse veto, never as a ranking signal.

\begin{algorithm}[t]
\caption{\method{} scoring and configuration selection}
\label{alg:daco}
\begin{algorithmic}[1]
\Require candidate configurations $\Gamma$; per machine: training normals
$\mathcal{B}=\mathcal{B}_\source\cup\mathcal{B}_\target$; labeled dev
machines
\Statex \textit{Scoring (per machine, per configuration
$\gamma\in\Gamma$):}
\State compute the configuration's training-bank calibration scores (LOO)
\State construct $\gamma$'s score transform: per-domain quantile maps
with shrinkage $m$ (\ref{eq:shrink}), per-domain $z$-score or
median-ratio, LDN, or the identity (raw)
\State score a test clip $x$ by applying the transform,
$\tilde A(x)=\sum_{g} w_g(x)\,c_g(A(x))$ for per-domain maps with
soft or hard weights $w_g$
\Statex \textit{Selection:}
\For{$\gamma\in\Gamma$}
  \State $C(\gamma)\gets$ CV domain-balance criterion on training normals
  \hfill (\ref{eq:criterion})
\EndFor
\State $\mathcal{V}\gets\Gamma$ without its bottom decile by dev $\Omega$
\hfill (veto)
\State \Return $\arg\min_{\gamma\in\mathcal{V}} C(\gamma)$
\end{algorithmic}
\end{algorithm}

\section{Experimental Setup}\label{sec:setup}
\textbf{Systems.} Frozen backbones: BEATs
iter3+~AS2M~\cite{chen2022beats} (primary; time-mean-pooled frame features,
768-d), EAT-base~\cite{chen2024eat} (CLS utterance token, 768-d; checkpoint
\texttt{worstchan/EAT-base\_epoch30\_pretrain}), and PANNs
CNN14-16k~\cite{kong2020panns} (penultimate embedding, 2048-d; checkpoint
\texttt{Cnn14\_16k\_mAP=0.438}). Base scorer: kNN cosine distance,
$k\in\{1,2,4\}$. Configuration grid: raw; per-domain quantile calibration
(\cref{eq:shrink}) with $m\in\{0,10,30,100,300\}$ $\times$ assignment
$\{$soft, hard$\}$; per-domain $z$-score (the normalization family used by
GenRep~\cite{genrep2025}); per-domain median-ratio; and the local-density
normalization family (ratio/difference, density neighborhood
$K\in\{1,16\}$~\cite{wilkinghoff2025icassp,wilkinghoff2025localdensity},
reimplemented from the published equations---the AGPL reference
implementation was not consulted) plus calibration stacked on it. In total
45 configurations for the transfer study and 51 per backbone for the
ablations.

\textbf{Experiments.} E1 reproduces baselines; E2 maps the balance frontier
(dev set); E3 is the transfer and selection study; E4 repeats it across
backbones; E5 is the head-to-head with local-density normalization; E6
reports cost; E7 replicates the full pipeline on DCASE 2023 and 2024
(datasets and post-challenge ground truth from the respective official
Zenodo records and evaluator repositories) and pre-registers the DCASE
2026 forward test. The criterion uses $S{=}10$ splits, seed 0 (stability
over seeds 0--9 reported); all other components are deterministic.

\textbf{Hygiene.} All selection signals are computed from training normals
plus (for the veto only) labeled dev data. By construction, test labels and
domains of the evaluation machines enter only the metric module and one
explicitly-labeled oracle diagnostic; this can be verified directly in the
released code, whose data loaders inject ground truth exclusively into test
clips and whose selection paths consume training embeddings alone. The
metric implementation passes a released 2000-trial randomized differential
test against the official evaluator's per-domain filtering logic (maximum
absolute deviation $0$), and headline results were re-scored end-to-end by
the official evaluator on exported per-clip scores---which we release for
every configuration appearing in a table, together with a configuration
manifest, checkpoint SHA-256 pins, the environment lock, and scripts for
every statistic in this paper, including the bootstrap and permutation
procedures.

\section{Results}\label{sec:results}

\subsection{Baseline reproduction (E1)}\label{sec:results-e1}
BEATs + kNN ($k{=}1$) attains dev $\Omega=58.73$ versus $56.26$ (Simple AE)
and $55.34$ (Selective Mahalanobis) for the official baselines (dev-set
aggregates recomputed as harmonic means of the organizers' published
per-machine values); our frozen-embedding Selective-Mahalanobis and GMM
backends reach $55.83$ and $52.03$, supporting kNN as the base scorer. The
raw source/target gap is large (mean $\auc_\source$ $70.8$ vs.\
$\auc_\target$ $57.0$), reproducing the imbalance the protocol is designed
to expose.

\subsection{The balance frontier (E2)}\label{sec:results-frontier}
\begin{figure}[t]
\centering
\includegraphics[width=\columnwidth]{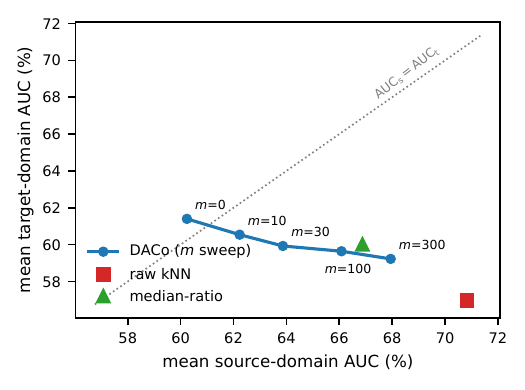}
\caption{The source/target balance frontier: mean source vs.\ target AUC
over the seven dev machines (BEATs backbone; soft assignment, $k{=}1$).
The prior strength $m$
(\cref{eq:shrink}) interpolates between the raw kNN ranking and full
per-domain equalization; the dotted line marks
$\auc_\source{=}\auc_\target$; the median-ratio variant sits near the
frontier.}
\label{fig:frontier}
\end{figure}

\Cref{fig:frontier} shows mean source vs.\ target AUC as $m$ sweeps from $0$
to $300$. The \emph{signed mean} gap moves smoothly from $+13.8$ points
(raw) through $-1.2$ at $m{=}0$ (a slight target overshoot, visible above
the diagonal in \cref{fig:frontier}); the median per-machine gap drops from
$15.2$ to $3.6$. Aggregate balance is thus controllable. Per-machine balance
is not the same thing: the mean \emph{absolute} gap only moves from $15.1$
to $14.0$, dominated by one machine (\texttt{ToyCar}) where calibration
overshoots violently (source $75.8\to35.2$, target $59.0\to84.5$).

\begin{figure}[t]
\centering
\includegraphics[width=\columnwidth]{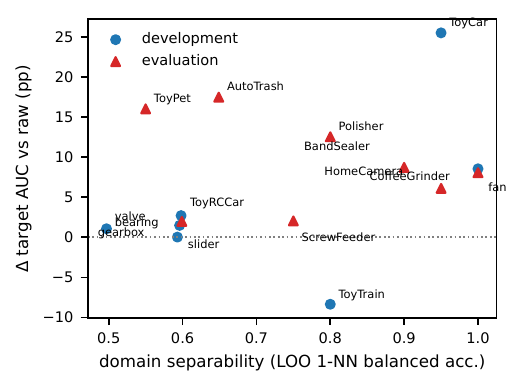}
\caption{Per-machine target-AUC gain of calibration ($m{=}0$, soft
assignment, $k{=}1$, BEATs) over raw kNN ($k{=}1$) versus domain
separability of the training normals (LOO 1-NN
balanced accuracy). Circles: dev machines; triangles: evaluation machines
(post-challenge ground truth used for metric computation only). The
structure--response relationship is clean on dev but weak overall
($\rho_s{=}{+}0.39$, $p{=}0.15$): \texttt{ToyPet} and \texttt{AutoTrash}
gain at low separability, \texttt{ToyTrain} loses at high separability.}
\label{fig:gainsep}
\end{figure}

The mechanism of that failure is diagnostic (released as
\texttt{diagnostics.csv}): on \texttt{ToyCar} only 6\% of anomalies exceed
the largest of the ten target LOO calibration scores while 99\% exceed the
source 90th percentile, so target-assigned anomalies are pulled below
mid-ranked source normals---even with oracle domain assignment. On the dev
set, gains and losses follow Stage-1 assignment accuracy (measured on test
normals; also released): \texttt{fan} (100\%) gains $+8.5$ target-AUC points
at $m{=}0$, \texttt{ToyCar} (85\%) gains $+25.5$ on target while collapsing
on source, and chance-level machines
(\texttt{bearing}/\texttt{slider}/\texttt{valve}, 51--54\%) are unaffected.
We caution that this clean structure--response picture is a \emph{dev-set
observation}: across all 15 machines the rank correlation between train-side
domain separability and target-AUC gain is only $+0.39$ ($p{=}0.15$), with
low-separability gainers (\texttt{ToyPet} $+16.0$ at $0.55$;
\texttt{AutoTrash} $+17.5$ at $0.65$) and one separable loser
(\texttt{ToyTrain} $-8.4$ at $0.80$); see \cref{fig:gainsep}. Dev $\Omega$
consequently does \emph{not} improve along the frontier (raw $58.73$;
$m{=}0$: $56.29$)---which is precisely why calibration strength must be
selected, not fixed.

Finally, the operating-point property itself: at matched label-free global
thresholds (nominal levels 0.1--0.5), calibration approximately halves the
mean per-machine cross-domain FPR imbalance relative to raw scoring when
averaged over all 15 machines
($|\mathrm{FPR}_\source-\mathrm{FPR}_\target|$ $0.16$ vs.\ $0.31$ at the
$0.1$ level; on the seven dev machines alone the imbalance instead
increases, driven by \texttt{ToyCar}), although absolute FPR levels exceed
nominal under the train$\to$test session shift ($\approx0.21$ at nominal
$0.10$)---the approximate-equalization framing of \cref{sec:stage2},
validated and bounded.

\subsection{Transfer and label-free selection (E3)}
\label{sec:results-transfer}
\begin{figure*}[t]
\centering
\includegraphics[width=0.92\textwidth]{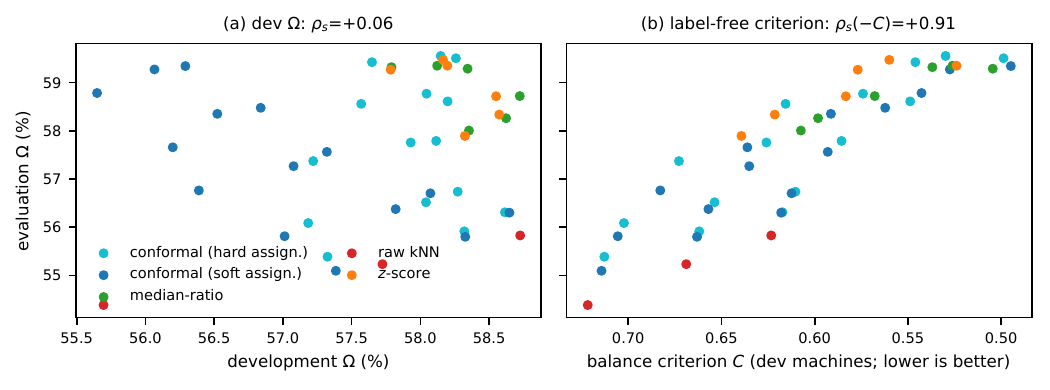}
\caption{What predicts evaluation-set performance? Each point is one of 45
configurations (BEATs backbone); colors denote the five method groups
(the soft and hard variants of median-ratio and $z$-score share their
group's color, so the seven families map to five colors). (a)~Dev
$\Omega$ is uninformative---the standard selection signal fails.
(b)~The label-free CV domain-balance criterion $C$ (\cref{eq:criterion},
computed on dev machines; axis reversed so better is rightward)
rank-predicts evaluation $\Omega$.}
\label{fig:transfer}
\end{figure*}

\Cref{fig:transfer} is the central result. Across the 45-configuration grid,
dev $\Omega$ carries no rank information about evaluation $\Omega$
($\rho_s{=}{+}0.06$, family-block bootstrap 95\% CI $[-0.39,+0.31]$,
panel a); the label-free criterion computed on dev machines predicts it at
$\rho_s{=}{+}0.91$ (panel b). The disconnect is structural, not noise: dev
and evaluation $\Omega$ respond to the calibration-strength knob in opposite
directions ($\rho_s$ with $\log m$: $+0.45$ vs.\ $-0.96$; within the
conformal-soft family the dev--eval rank correlation is $-0.79$). The dev
range is also compressed (3.1 vs.\ 5.2 points), so dev selection is
noise-limited on top of being directionally wrong.

Because the 45 configurations cluster into seven method families (raw and
the soft/hard variants of conformal, median-ratio, and $z$-score, each
family pooling $k$ and, for conformal, $m$), we report
clustered uncertainty computed by released scripts: a family-block bootstrap
(resampling the seven families, 5000 draws) gives a 95\% CI of
$[+0.83,+0.95]$; the rank correlation of the seven family means is $+0.93$
(exact permutation test over all $7!$ orderings: one-sided $p{=}0.0034$,
two-sided $p{=}0.0067$); a machine-level bootstrap (resampling the 7 dev
machines for the criterion and the 8 evaluation machines for $\Omega$) is
much wider, $[+0.24,+0.97]$, reflecting the small machine sets. The
correlation survives family-structure stress tests: conformal-only $+0.89$,
leave-one-family-out $+0.90$ to $+0.92$, dropping all conformal
configurations $+0.92$. On the extended 51-configuration grid (adding the
LDN family) the same correlation is $+0.81$ with family-block CI
$[+0.43,+0.93]$---the grid used for all cross-year comparisons below.

\emph{Is the correlation just re-reading $m$?} Within conformal families the
criterion is close to monotone in $m$ by construction, so we deconfound
explicitly: partialling out $\log(m{+}1)$ and family membership across the
full grid leaves a residual association of $+0.71$ (partial Spearman,
$p{<}10^{-6}$)---the criterion carries information beyond the
calibration-strength trend, chiefly by ordering the method families. Within
the 30 conformal configurations alone, after controlling $m$, assignment,
and $k$, the residual signal is weak ($+0.21$, $p{=}0.27$): consistent with
the region-selector reading, the criterion distinguishes families and
calibration regimes, not neighbors within a family.

\begin{table}[t]
\centering
\caption{Model selection on DCASE 2025 Task 2 (evaluation set, official
$\Omega$, \%; BEATs backbone, 45-configuration grid). Rows 2--4 select with
the label-free criterion (\cref{eq:criterion}); row 1 uses labeled dev data
(standard practice). ``$\pm$'' is the standard deviation over ten
criterion seeds (pool of seven $k{=}1$ soft configurations from the
45-grid; the nine-configuration 51-grid pool of \cref{tab:backbones}
gives $59.63$). Evaluation ground truth enters metric computation only.
Official-evaluator-verified.}
\label{tab:selection}
\setlength{\tabcolsep}{4pt}
\begin{tabular}{llc}
\toprule
Selection rule & Chosen configuration & Eval $\Omega$ \\
\midrule
dev $\Omega$ (standard) & raw kNN ($k{=}1$) & 55.83 \\
criterion on dev machines & conf.\ soft $m{=}0$, $k{=}1$ & \textbf{59.34} \\
criterion on eval train normals & conf.\ soft $m{=}0$, $k{=}2$ & 59.27 \\
per-machine blind (10 seeds) & mixed & $59.14\pm0.12$ \\
\midrule
oracle best fixed config & conf.\ hard $m{=}0$, $k{=}2$ & 59.55 \\
\bottomrule
\end{tabular}
\end{table}

\Cref{tab:selection} translates the correlation into the deliverable on this
grid. Selecting the configuration that minimizes the criterion on the
\emph{dev} machines---selection that uses no evaluation-machine data at
all---and deploying
it unchanged yields $\Omega=59.34$ vs.\ $55.83$ for the best-dev-$\Omega$
configuration: $+3.5$ points from changing only the selection yardstick
(machine-level jackknife 95\% CI $[2.2,4.8]$, leave-one-machine-out range
$[3.2,4.0]$; paired clip-level stratified bootstrap $[2.5,4.7]$),
within $0.21$ of the best fixed configuration in hindsight. The two smallest
criterion values differ by $0.004$---inside seed noise---so the pick can
flip to conf.\ hard $m{=}0$, $k{=}1$ ($59.50$); all plausible picks lie in
$59.3$--$59.6$. Per-machine blind selection over a fixed nine-configuration pool (the
$k{=}1$ soft-assignment representatives: raw, the five conformal
strengths, median-ratio, LDN-ratio $K{=}1$, and calibration-on-LDN) attains $59.14\pm0.12$ over ten criterion seeds (min
$58.91$). Within the top cluster the criterion's ranking is at chance---it
is a region selector, and the whole low-criterion region is good.

For context on the absolute scale: the official evaluation-set baselines
score $56.51$ (Selective Mahalanobis) and $54.43$ (Simple AE), and the top
three of 35 teams scored $61.63$, $61.57$, $61.20$~\cite{dcase2025task2}.
The dev-selected raw system ($55.83$) falls \emph{below} the stronger
official baseline---rank 23 among teams---despite beating both baselines on
dev (\cref{sec:results-e1}): dev-based selection does not merely lose
points, it publishes a system worse than the baseline. Criterion-based
selection turns the identical pool into rank~7 ($59.34$), and with the E5
family into rank~4 ($61.05$, \cref{sec:results-ablations})---as a single
training-free system with no ensembling, no fine-tuning, and post-challenge
timing.

\subsection{Robustness across backbones (E4)}\label{sec:results-ablations}
\Cref{tab:backbones} repeats the study on the extended 51-configuration grid
for three frozen backbones. The criterion (computed on dev machines) remains
a reliable rank predictor on every backbone ($\rho_s$ $+0.81$/$+0.80$/$+0.67$),
whereas dev $\Omega$ swings from useless ($-0.03$, BEATs; $+0.11$, EAT) to
strong ($+0.84$, PANNs): with PANNs embeddings, dev and evaluation machines
happen to agree, so standard selection works \emph{there}---but one cannot
know in advance which regime a given system is in, and only the criterion is
reliable in both. Guarded selection (bottom-decile veto + criterion) gains
$+5.2$ (BEATs) and $+2.3$ (EAT) points over dev-$\Omega$ selection and costs
$-1.2$ on PANNs, where dev selection is coincidentally oracle-optimal; the
mean gain is $+2.1$. \Cref{tab:backbones} also situates selection against
reference rules: a uniformly random configuration averages
$54.8$--$57.8$, and the fixed a-priori default (full equalization,
conf.\ soft $m{=}0$, $k{=}1$) is itself a strong label-free rule
($59.34$/$58.87$/$55.55$)---guarded selection beats it only on BEATs
($+1.7$, where it finds the LDN stack) and ties it elsewhere, so the
criterion's marginal value over a sensible fixed default is concentrated
where the grid contains a genuinely better family. The unguarded
criterion-only row makes the veto's contribution explicit
($45.26$ on EAT without it). On the grid itself, soft vs.\ hard assignment
and $k\in\{1,2,4\}$ shift results by well under a point; the calibration
variant matters less than the strength $m$.

\emph{Veto disclosure.} The veto was introduced \emph{after} observing that
criterion-only selection picks a degenerate configuration on EAT
(\cref{sec:results-ldn}); it is not a pre-registered rule, and we report it
transparently. The threshold form ($\delta$ from dev-best) selects
identically for $\delta\in[3,8]$ on BEATs and PANNs; on EAT, $\delta\le5$
also excludes the three best evaluation configurations (the LDN family,
whose dev deficit is ${\approx}5.5$ points there), costing $0.9$ points
against the veto-free oracle, while $\delta{=}8$ recovers them ($59.75$).
The rank-based bottom-decile form used in \cref{tab:backbones} makes the
same picks as $\delta{=}5$ on all three backbones. The residual
meta-selection risk (choosing the veto form) is bounded by this spread
($58.87$--$59.75$ on EAT) and is disclosed rather than optimized away.

\begin{table}[t]
\centering
\caption{Selection across frozen backbones (evaluation-set official
$\Omega$, \%; 51 configurations per backbone). ``Guarded criterion'' =
bottom-decile dev-$\Omega$ veto + criterion minimum (\cref{alg:daco}).
Reference rows: expected score of a uniformly random configuration
(mean$\pm$sd over the 51 configurations), the fixed a-priori default
(full equalization: conf.\ soft $m{=}0$, $k{=}1$; designated before the
E7 replication), and the unguarded criterion minimum. Best non-oracle
rule per backbone in bold.}
\label{tab:backbones}
\setlength{\tabcolsep}{4pt}
\begin{tabular}{lccc}
\toprule
 & BEATs & EAT & PANNs \\
\midrule
random configuration & $57.8{\pm}2.1$ & $57.5{\pm}2.1$ & $54.8{\pm}1.5$ \\
raw kNN ($k{=}1$) & 55.83 & 56.59 & 54.23 \\
fixed default (conf.\ soft $m{=}0$) & 59.34 & 58.87 & 55.55 \\
selected by dev $\Omega$ & 55.83 & 56.59 & \textbf{56.83} \\
criterion only (no veto) & 61.05 & 45.26 & 55.58 \\
selected by guarded criterion & \textbf{61.05} & \textbf{58.87} & 55.58 \\
per-machine blind (fixed pool) & 59.63 & 60.38 & 56.66 \\
\midrule
oracle best fixed config & 61.05 & 59.76 & 56.83 \\
\midrule
$\rho_s$(dev $\Omega$, eval $\Omega$) & $-0.03$ & $+0.11$ & $+0.84$ \\
$\rho_s$($-$criterion, eval $\Omega$) & $+0.81$ & $+0.80$ & $+0.67$ \\
\bottomrule
\end{tabular}
\end{table}

\subsection{Comparison with local-density normalization (E5)}
\label{sec:results-ldn}
\Cref{tab:ldn} compares against our reimplementation of local-density
normalization (LDN)~\cite{wilkinghoff2025icassp,wilkinghoff2025localdensity}
on identical embeddings; all LDN statements in this paper concern this
reimplementation from the published equations, not the authors' system
(which couples the normalization to trained embeddings). The density
neighborhoods $K\in\{1,16\}$ cover both papers' recommendations: $K{=}1$
for the official metric~\cite{wilkinghoff2025localdensity} and $K{=}16$
from the ICASSP study~\cite{wilkinghoff2025icassp}. Three observations.
First, LDN is strong on the evaluation set (ratio variant: $61.04$ for
both $K$; the agreement is coincidental at two decimals,
$61.039$ vs.\ $61.040$---the density is computed over the full
1000-clip bank, so $K{=}16$ is well-defined despite the 10-clip target
subset)---but \emph{weaker than raw on the dev set} ($56.5$/$57.2$ vs.\
$58.7$), so dev-based selection would reject it: even a strong normalization
method is a victim of the selection trap this paper addresses, and the
criterion selects into its family. Second, calibration stacked on LDN scores
($61.05$) is statistically indistinguishable from LDN alone ($61.04$; the
margin is far inside seed noise)---the value of the stack is that it is what
the guarded criterion picks, not that composition adds points. Third, the
difference variant with a large density neighborhood (diff, $K{=}16$)
collapses to a degenerate, well-balanced but non-detecting score map
($44.8$--$49.7$ evaluation $\Omega$ across backbones). This is expected by
construction---subtracting a large summed-distance density term drowns the
query distance---and is consistent with the original authors' report that,
under the official metric, ``performance drops rapidly when increasing the
value of $K$''~\cite{wilkinghoff2025localdensity} (the difference variant
itself is published only at $K{=}1$, so $K{=}16$-difference is unanchored
in prior work); it is exactly the class of configuration a balance-only
criterion cannot reject. Criterion-only selection falls for it in three of
the five year--backbone settings where the family is present---both
replication years and the EAT backbone in 2025
(\cref{tab:backbones,tab:years})---the concrete failure that motivates the
viability veto.

\emph{Reimplementation validation.} The journal version
of~\cite{wilkinghoff2025localdensity} reports frozen-BEATs LDN results on
DCASE 2023 and 2024 under the official metric, which our replication years
cover with the same datasets: their raw-kNN baselines ($59.0$ and $58.2$
evaluation $\Omega$) are within $1.1$ points of ours ($57.95$, $57.64$),
and the qualitative pattern of their $K{=}1$ gains---evaluation-set gains
much larger than dev-set gains, 2023 gains larger than 2024---reproduces
in our reimplementation (e.g., ratio $K{=}1$ on evaluation:
$+4.6$/$+1.5$ vs.\ their $+8.6$/$+4.2$). Absolute deltas are roughly
halved, consistent with the different embedding readout (they use a
6144-d flattened patch embedding scored with MSE; we use the 768-d
temporal mean with cosine distance); exact reproduction is not possible
from the papers alone, and all LDN rows here should be read as our
reimplementation under our readout.

\begin{table}[t]
\centering
\caption{\method vs.\ local-density normalization on identical BEATs
embeddings (official $\Omega$, \%). LDN reimplemented from the equations
of~\cite{wilkinghoff2025icassp,wilkinghoff2025localdensity}. The dev/eval
agreement in the median-ratio row is coincidental ($58.7233$/$58.7189$).}
\label{tab:ldn}
\begin{tabular}{lcc}
\toprule
System & Dev $\Omega$ & Eval $\Omega$ \\
\midrule
raw kNN ($k{=}1$) & \textbf{58.73} & 55.83 \\
LDN ratio, $K{=}1$ & 56.47 & 61.04 \\
LDN ratio, $K{=}16$ & 57.24 & 61.04 \\
LDN difference, $K{=}1$ & 57.94 & 60.23 \\
LDN difference, $K{=}16$ (degenerate) & 50.39 & 49.68 \\
per-domain $z$-score (soft, $k{=}1$) & 58.55 & 58.71 \\
per-domain median-ratio (soft, $k{=}1$) & 58.72 & 58.72 \\
\method (conf.\ soft $m{=}0$, $k{=}1$) & 56.29 & 59.34 \\
\method on LDN-ratio $K{=}1$ & 56.56 & \textbf{61.05} \\
\bottomrule
\end{tabular}
\end{table}

\subsection{Replication across challenge years and a pre-registered
forward test (E7)}\label{sec:results-years}
The 2025 study is one dev$\to$eval machine-set pair. We replicated the
complete pipeline---identical 51-configuration grid, criterion, veto, and
BEATs backbone---on DCASE 2023 (7 evaluation machine types) and DCASE 2024
(9 types), scoring with the respective official post-challenge ground
truth. \Cref{tab:years} shows the result. Two findings replicate without
qualification: dev $\Omega$ carries no rank information about evaluation
$\Omega$ in \emph{any} year ($\rho_s{=}{+}0.18/{+}0.13/{-}0.03$, all
n.s.\ under every treatment we applied); and the degenerate
LDN-diff-$K{=}16$ trap is selected by the unguarded criterion in both
replication years ($47.6/47.5$)---and on the EAT backbone in 2025
(\cref{tab:backbones})---that is, in three of the five year--backbone
settings containing the family, so the veto is not a 2025 artifact, and
the guarded rule never selects a broken configuration.

The criterion's predictive power requires more care. Raw full-grid
correlations are $+0.54$ (2023), $-0.10$ (2024), $+0.81$ (2025), but the
naive per-configuration $p$-values overstate evidence because
configurations cluster into families: under the same family-block
apparatus used in \cref{sec:results-transfer} (block bootstrap over the
ten families of the 51-grid; Monte-Carlo permutation of family means,
200k draws), only 2025 survives---CI $[+0.43,+0.93]$ excluding zero---
while the 2023 CI crosses zero ($[-0.09,+0.70]$; family-mean permutation
$p{=}0.23$) and, after partialling out family membership and $\log m$,
the residual 2023 association is in fact \emph{negative} ($-0.61$). We
therefore claim demonstrated criterion transfer in \emph{one year of
three} and read 2023 as family-composition-driven. Selection deltas tell
the same story (machine-level jackknife CIs): guarded selection beats
dev-$\Omega$ selection significantly only in 2025 ($+5.2$,
$[1.3,9.2]$; 2023 $+1.3$ $[-1.2,+3.8]$; 2024 $-1.4$ $[-4.0,+1.3]$),
and its differences against the fixed full-equalization default are
within machine-level noise in all three years ($-0.2$ $[-1.7,+1.4]$;
$-1.4$ $[-3.0,+0.2]$; $+1.7$ $[-1.2,+4.6]$). The fixed default
(conf.\ soft $m{=}0$, $k{=}1$)---designated \emph{before} the
replication was run, as the E3 criterion pick and the frontier's
full-equalization endpoint---matches or beats guarded selection in both
replication years and is the honest recommendation when nothing is known
about the target year; per-machine blind selection tracks it closely
($63.38/57.63/59.63$, \cref{tab:years}), as its per-machine choices are
dominated by $m{=}0$ variants. The criterion's demonstrated added value
is concentrated in 2025, where the configuration space contains
family-level structure worth finding (the LDN stack); the veto's value
--- catching degenerate configurations --- replicates in every year where
the degenerate family is present.

\begin{table}[t]
\centering
\caption{Replication across challenge years (BEATs, 51 configurations,
evaluation-set official $\Omega$, \%). Correlation significance is
assessed by family-block bootstrap (ten families, 5000 draws);
``n.s.''\ = the 95\% CI crosses zero. $^{\dagger}$naive
per-configuration $p{<}10^{-4}$, but the family-block CI
$[-0.09,+0.70]$ crosses zero and the family-mean permutation
$p{=}0.23$; see \cref{sec:results-years}. Per-machine blind uses the
nine-configuration pool of \cref{tab:backbones}, criterion seed 0.
Best label-free rule per year in bold.}
\label{tab:years}
\setlength{\tabcolsep}{4pt}
\begin{tabular}{lccc}
\toprule
 & 2023 & 2024 & 2025 \\
\midrule
$\rho_s$(dev $\Omega$, eval $\Omega$) & $+0.18$ n.s. & $+0.13$ n.s. & $-0.03$ n.s. \\
$\rho_s$($-$criterion, eval $\Omega$) & $+0.54^{\dagger}$ & $-0.10$ n.s. & $+0.81$ \\
\midrule
raw kNN ($k{=}1$) & 57.95 & 57.64 & 55.83 \\
selected by dev $\Omega$ & 62.23 & 57.64 & 55.83 \\
criterion only (no veto) & 47.60 & 47.48 & 61.05 \\
selected by guarded criterion & 63.52 & 56.29 & 61.05 \\
fixed default (conf.\ soft $m{=}0$) & \textbf{63.68} & \textbf{57.69} & 59.34 \\
per-machine blind (fixed pool) & 63.38 & 57.63 & 59.63 \\
\midrule
oracle best fixed config & 64.14 & 60.11 & 61.05 \\
\bottomrule
\end{tabular}
\end{table}

\emph{Frozen forward test on DCASE 2026.} DCASE 2026 retains the
source/target single-threshold protocol with two-channel, noise-focused
recordings~\cite{dcase2026task2}. Its timeline at the time of writing:
the evaluation audio was published on 2026-06-01 (Zenodo record
20437238), challenge submissions closed on 2026-06-15, and the team
leaderboard was announced on 2026-06-30 (175 submissions; top official
score $70.24$; official MSE baseline $59.80$)---but the \emph{per-clip
evaluation ground truth} that offline scoring requires has not been
released. We ran the full selection on the released 2026
\emph{development} set and froze the outcome in the public repository
(\texttt{PREREGISTRATION.md}: selected configuration per channel policy,
veto rule, criterion seeds, and SHA-256 hashes of the selection
artifacts) on 2026-07-03---after the
evaluation audio and leaderboard were public, but before the per-clip
ground truth became available---so the frozen configuration could not
have been evaluated, directly or indirectly, at freezing time.
The primary frozen pick (channel~0 policy) is the per-domain median-ratio
configuration with hard assignment ($k{=}2$)---notably \emph{not} the
fixed default, making the test discriminative. We commit to scoring the
frozen pick, the dev-$\Omega$ pick, and the fixed default with the
official evaluator as soon as the ground truth is published, and to
reporting the outcome in this section whatever it is.

\subsection{Computational cost (E6)}\label{sec:results-efficiency}
All results were produced on one consumer GPU (RTX 4070~Ti SUPER, 16\,GB).
Embedding extraction dominates and runs once (\cref{tab:efficiency}); on
cached embeddings a full per-machine configuration cycle (base scores, LOO,
calibration, and the 10-split criterion) takes ${\sim}0.15$\,s on CPU, so
the complete 51-configuration $\times$ 15-machine study runs in under four
minutes per backbone. Peak extraction VRAM is below 1\,GB.

\begin{table}[t]
\centering
\caption{Efficiency (single RTX 4070 Ti SUPER; 10-s clips at 16\,kHz,
batch 8).}
\label{tab:efficiency}
\begin{tabular}{lcccc}
\toprule
Backbone & Params (M) & Dim & ms/clip & VRAM (MB) \\
\midrule
BEATs & 90.3 & 768 & 9.4 & 824 \\
EAT-base & 90.0 & 768 & 7.2 & 621 \\
PANNs CNN14 & 81.0 & 2048 & 1.9 & 703 \\
\bottomrule
\end{tabular}
\end{table}

\section{Discussion and Limitations}\label{sec:discussion}
\paragraph{Why does the criterion transfer while dev $\Omega$ does not?}
The criterion measures how well a configuration equalizes the two domains'
normal-score distributions---a property of the \emph{operating point}, not
of any machine's anomaly geometry. The evaluation machines reward balance
(raw target AUCs as low as $33.7$ on \texttt{CoffeeGrinder}; per-machine
table in the Appendix); several dev machines punish it
(\cref{sec:results-frontier}). A signal tied to threshold placement rather
than to dev-set anomaly idiosyncrasies tracks the quantity that generalizes.

\paragraph{Scope of the transfer evidence}
On dev machines the criterion does \emph{not} predict dev $\Omega$
($\rho_s{=}{+}0.16$); blind selection there would have cost $2.7$ points.
Our claim is not that the criterion predicts any set's performance, but
that it predicts \emph{evaluation} performance where dev $\Omega$ does
not---and the three-year replication (\cref{sec:results-years}) bounds
this claim honestly. What replicates without exception across
2023/2024/2025 and across three backbones: dev $\Omega$ is never a
\emph{reliable} selection signal---it swings from useless to
coincidentally optimal (PANNs; the 2023 dev pick gained $+4.3$ over raw)
and one cannot know in advance which regime applies---and a well-balanced
degenerate configuration always exists in the pool (the unguarded
criterion selects it in three of five year--backbone settings). What is
year-dependent: the criterion's predictive power ($+0.54$, $-0.10$,
$+0.81$ raw; only 2025 robust to family clustering), which tracks how
much family-level structure the configuration space contains and how much
headroom the year offers (the 2024 oracle is only $2.5$ points above
raw). A practitioner cannot tell the regimes apart in advance; the
defensible default is therefore the fixed full-equalization
configuration, with criterion-based selection as the no-worse-than
upgrade whenever a heterogeneous configuration pool is on the table.
The added noise axis of DCASE 2026 is itself a new confound for the
criterion---the pre-registered forward test
(\cref{sec:results-years}) will answer it prospectively.

\paragraph{Balance is necessary, not sufficient}
A label-free criterion cannot observe detection power, and degenerate score
maps achieve excellent balance while detecting nothing
(\cref{sec:results-ldn}). The viability veto closes this with labeled dev
data used at the coarse level where they are reliable. This division of
labor---dev labels veto broken configurations, the label-free criterion
ranks the viable ones---is, we believe, the right general template for
model selection under the first-shot protocol; its residual meta-choices are
quantified in \cref{sec:results-ablations}.

\paragraph{Granularity and validity of ten-clip calibration}
With $n_\target{=}10$ the quantile map has data support only down to
exceedance $1/11\approx0.091$; nearly all of the $\pauc$ range is
extrapolated, and pAUC additionally pools domains, so the support boundary
maps onto the metric only loosely. Empirically the maps \emph{undercover}
under train$\to$test session shift: the realized false-positive rate exceeds
the nominal level (${\approx}0.21$ at nominal $0.10$,
\cref{sec:results-frontier}), i.e., the calibrated threshold is
anti-conservative in absolute terms even where cross-domain balance
improves. The
weighted-conformal small-$n$ analysis of~\cite{weightedconformal2026} and
group-conditional constructions~\cite{ding2023classconditional} suggest
principled refinements.

\paragraph{Criterion scale across families}
The criterion's absolute value is not perfectly comparable across method
families (hard-assignment configurations outperform soft ones by up to
${\approx}2.3$ $\Omega$ points at closely matched criterion values
($|\Delta C|<0.005$); within-family it is
reliable. Cross-family selection should therefore be read as
region-selection, consistent with the $0.21$-point oracle gap on the E3
grid (the gaps are $0.9$ and $1.25$ on EAT and PANNs,
\cref{tab:backbones}).

\paragraph{Threats to validity}
(i)~Post-challenge ground truth could leak through experimenter iteration;
selection signals are computed from training normals (plus dev labels for
the veto), the pipeline was audited for leakage, and headline numbers were
reproduced by the official evaluator on exported scores. (ii)~The evaluation
set is scored offline; the evaluator is the official one. (iii)~Criterion
seeds shift composed selection by $\pm0.12$ $\Omega$; fixed-configuration
picks can flip within the $59.3$--$59.6$ cluster. (iv)~The veto was
introduced after observing a failure case and its meta-choices are
disclosed, not pre-registered.

\section{Conclusion}\label{sec:conclusion}
We reframed the two open problems of first-shot domain-generalized ASD---the
source/target trade-off and dev$\to$eval non-transfer---as calibration and
model-selection problems. A per-domain quantile-calibration layer with a
shrinkage knob makes calibration strength selectable; a label-free
cross-validated balance criterion, paired with a coarse dev-side viability
veto, selects the operating point. Across three challenge years,
development score is never a reliable selection signal and well-balanced
degenerate configurations always lurk in the pool; the criterion's
demonstrated transfer is concentrated in DCASE 2025, where it lifts a
training-free frozen-embedding system from below the official baseline to
fourth place among challenge teams at ${\sim}0.15$\,s per configuration
cycle---while in both replication years a fixed full-equalization default
matches or beats criterion-based selection. The method applies unchanged
on top of any frozen embedding and training-free backend; the frozen
DCASE 2026 forward test, fixed before the per-clip evaluation ground
truth became available, will extend the evidence prospectively.

\section*{Data and Code Availability}
\sloppy
The DCASE Task~2 datasets are public on Zenodo: 2025 (records 15097779,
15392814, 15519362), 2023 (7882613, 7830345, 7860847), 2024 (10902294,
11259435, 11363076), and the 2026 development set (19336329); 2023/2024
evaluation ground truth comes from the official
\texttt{nttcslab/dcase202\{3,4\}\_task2\_evaluator} repositories, and the
pre-registration of the 2026 forward test (frozen configuration, artifact
hashes) is \texttt{PREREGISTRATION.md} in our repository. The 2025
ground truth is published in the official evaluator repository
(\texttt{nttcslab/dcase2025\_task2\_evaluator}). All experiment code,
per-machine result CSVs, diagnostic artifacts, and the scripts computing
every statistic in this paper (bootstrap CIs, FPR coverage, criterion floor,
differential metric test) are available at
\url{https://github.com/polestvr/daco-experiments}. Backbone checkpoints:
BEATs iter3+~AS2M, EAT \texttt{worstchan/EAT-base\_epoch30\_pretrain} (both
MIT), PANNs \texttt{Cnn14\_16k\_mAP=0.438} (Zenodo record 3987831, CC-BY).
The local-density-normalization comparison is a clean-room reimplementation
from the published equations; the AGPL-licensed reference implementation was
not consulted.

\appendix
\section*{Per-Machine Results}
\Cref{tab:permachine} reports per-machine metrics (BEATs backbone) for the
dev-selected system (raw kNN), the E3 criterion pick (conf.\ soft $m{=}0$,
$k{=}1$), and the guarded pick (\method on LDN-ratio $K{=}1$).

\begin{table*}[t]
\centering
\caption{Per-machine results (BEATs; \%). For each system:
$\auc_\source$ / $\auc_\target$ / $\pauc$.}
\label{tab:permachine}
\small
\begin{tabular}{l ccc ccc ccc}
\toprule
 & \multicolumn{3}{c}{raw kNN ($k{=}1$)} &
   \multicolumn{3}{c}{conf.\ soft $m{=}0$, $k{=}1$} &
   \multicolumn{3}{c}{\method on LDN ($K{=}1$)} \\
\cmidrule(lr){2-4}\cmidrule(lr){5-7}\cmidrule(lr){8-10}
Machine & $\auc_\source$ & $\auc_\target$ & $\pauc$
        & $\auc_\source$ & $\auc_\target$ & $\pauc$
        & $\auc_\source$ & $\auc_\target$ & $\pauc$ \\
\midrule
\multicolumn{10}{l}{\emph{development machines}}\\
bearing & 60.6 & 48.0 & 57.5 & 60.3 & 49.1 & 57.6 & 67.1 & 64.2 & 62.5 \\
fan & 61.2 & 41.2 & 48.8 & 52.3 & 49.7 & 48.4 & 43.4 & 57.0 & 52.3 \\
gearbox & 65.5 & 51.2 & 55.0 & 61.2 & 53.9 & 55.0 & 63.4 & 58.0 & 54.6 \\
slider & 75.4 & 52.9 & 51.2 & 74.6 & 54.4 & 51.4 & 72.5 & 48.7 & 55.3 \\
ToyCar & 75.8 & 59.0 & 52.1 & 35.2 & 84.5 & 48.4 & 32.6 & 83.3 & 50.7 \\
ToyTrain & 85.0 & 69.8 & 53.8 & 65.0 & 61.4 & 52.4 & 67.7 & 63.8 & 49.8 \\
valve & 72.3 & 76.7 & 61.5 & 72.9 & 76.7 & 61.7 & 66.6 & 63.4 & 56.7 \\
\midrule
\multicolumn{10}{l}{\emph{evaluation machines}}\\
AutoTrash & 79.6 & 48.7 & 50.8 & 72.1 & 66.2 & 50.8 & 77.1 & 69.2 & 56.8 \\
BandSealer & 70.3 & 47.3 & 55.0 & 63.8 & 56.0 & 56.5 & 63.2 & 63.3 & 59.8 \\
CoffeeGrinder & 74.6 & 33.7 & 50.1 & 70.4 & 41.7 & 52.3 & 77.3 & 43.6 & 53.4 \\
HomeCamera & 74.3 & 34.8 & 51.7 & 69.0 & 40.9 & 51.8 & 65.6 & 48.0 & 52.3 \\
Polisher & 75.2 & 40.4 & 51.7 & 69.3 & 52.9 & 53.0 & 76.9 & 52.1 & 55.2 \\
ScrewFeeder & 80.1 & 88.7 & 72.0 & 81.9 & 90.7 & 75.3 & 83.8 & 94.8 & 78.0 \\
ToyPet & 80.0 & 47.1 & 56.6 & 68.0 & 63.1 & 57.9 & 68.9 & 72.7 & 58.7 \\
ToyRCCar & 63.4 & 56.6 & 52.3 & 61.0 & 58.6 & 52.4 & 58.8 & 44.2 & 51.6 \\
\midrule
\multicolumn{1}{l}{$\Omega$ (dev)} & \multicolumn{3}{c}{58.73}
 & \multicolumn{3}{c}{56.29} & \multicolumn{3}{c}{56.56} \\
\multicolumn{1}{l}{$\Omega$ (eval)} & \multicolumn{3}{c}{55.83}
 & \multicolumn{3}{c}{59.34} & \multicolumn{3}{c}{61.05} \\
\bottomrule
\end{tabular}
\end{table*}

\clearpage
\bibliographystyle{IEEEtran}
\bibliography{references}

\begin{thebibliography}{10}
\providecommand{\url}[1]{#1}
\csname url@samestyle\endcsname
\providecommand{\newblock}{\relax}
\providecommand{\bibinfo}[2]{#2}
\providecommand{\BIBentrySTDinterwordspacing}{\spaceskip=0pt\relax}
\providecommand{\BIBentryALTinterwordstretchfactor}{4}
\providecommand{\BIBentryALTinterwordspacing}{\spaceskip=\fontdimen2\font plus
\BIBentryALTinterwordstretchfactor\fontdimen3\font minus \fontdimen4\font\relax}
\providecommand{\BIBforeignlanguage}[2]{{%
\expandafter\ifx\csname l@#1\endcsname\relax
\typeout{** WARNING: IEEEtran.bst: No hyphenation pattern has been}%
\typeout{** loaded for the language `#1'. Using the pattern for}%
\typeout{** the default language instead.}%
\else
\language=\csname l@#1\endcsname
\fi
#2}}
\providecommand{\BIBdecl}{\relax}
\BIBdecl

\bibitem{dcase2025task2}
T.~Nishida, N.~Harada, D.~Niizumi, D.~Albertini, R.~Sannino, S.~Pradolini, F.~Augusti, K.~Imoto, K.~Dohi, H.~Purohit, T.~Endo, and Y.~Kawaguchi, ``Description and discussion on {DCASE} 2025 challenge task 2: First-shot unsupervised anomalous sound detection for machine condition monitoring,'' in \emph{Proc. Detection and Classification of Acoustic Scenes and Events Workshop (DCASE)}, 2025, pp. 55--59, official DCASE 2025 Task 2 organizers' paper.

\bibitem{harada2023firstshot}
N.~Harada, D.~Niizumi, Y.~Ohishi, D.~Takeuchi, and M.~Yasuda, ``First-shot anomaly sound detection for machine condition monitoring: A domain generalization baseline,'' in \emph{Proc. European Signal Processing Conference (EUSIPCO)}, 2023, pp. 191--195.

\bibitem{wilkinghoff2025localdensity}
K.~Wilkinghoff, H.~Yang, J.~Ebbers, F.~G. Germain, G.~Wichern, and J.~Le~Roux, ``Local density-based anomaly score normalization for domain generalization,'' \emph{IEEE Transactions on Audio, Speech and Language Processing}, vol.~33, pp. 4642--4652, 2025.

\bibitem{wilkinghoff2021subcluster}
K.~Wilkinghoff, ``Sub-cluster {AdaCos}: Learning representations for anomalous sound detection,'' in \emph{Proc. International Joint Conference on Neural Networks (IJCNN)}, 2021, pp. 1--8.

\bibitem{wilkinghoff2023featex}
------, ``Self-supervised learning for anomalous sound detection,'' in \emph{Proc. IEEE International Conference on Acoustics, Speech and Signal Processing (ICASSP)}, 2024, pp. 276--280.

\bibitem{wilkinghoff2024adaproj}
------, ``{AdaProj}: Adaptively scaled angular margin subspace projections for anomalous sound detection with auxiliary classification tasks,'' in \emph{Proc. Detection and Classification of Acoustic Scenes and Events (DCASE) Workshop}, 2024, pp. 186--190.

\bibitem{wilkinghoff2025review}
K.~Wilkinghoff, T.~Fujimura, K.~Imoto, J.~Le~Roux, Z.-H. Tan, and T.~Toda, ``Handling domain shifts for anomalous sound detection: A review of {DCASE}-related work,'' in \emph{Proc. Detection and Classification of Acoustic Scenes and Events (DCASE) Workshop}, 2025, pp. 20--24.

\bibitem{chen2022beats}
S.~Chen, Y.~Wu, C.~Wang, S.~Liu, D.~Tompkins, Z.~Chen, W.~Che, X.~Yu, and F.~Wei, ``{BEATs}: Audio pre-training with acoustic tokenizers,'' in \emph{Proc. International Conference on Machine Learning (ICML)}, 2023, pp. 5178--5193.

\bibitem{chen2024eat}
W.~Chen, Y.~Liang, Z.~Ma, Z.~Zheng, and X.~Chen, ``{EAT}: Self-supervised pre-training with efficient audio transformer,'' in \emph{Proc. International Joint Conference on Artificial Intelligence (IJCAI)}, 2024, pp. 3807--3815.

\bibitem{kong2020panns}
Q.~Kong, Y.~Cao, T.~Iqbal, Y.~Wang, W.~Wang, and M.~D. Plumbley, ``{PANNs}: Large-scale pretrained audio neural networks for audio pattern recognition,'' \emph{IEEE/ACM Transactions on Audio, Speech, and Language Processing}, vol.~28, pp. 2880--2894, 2020.

\bibitem{genrep2025}
P.~Saengthong and T.~Shinozaki, ``Deep generic representations for domain-generalized anomalous sound detection,'' in \emph{Proc. IEEE International Conference on Acoustics, Speech and Signal Processing (ICASSP)}, 2025.

\bibitem{zhou2026backends}
J.~Zhou and M.~Wang, ``Scoring backends matter more than pooling: A systematic study of training-free anomalous sound detection under domain shift,'' 2026.

\bibitem{wilkinghoff2025icassp}
K.~Wilkinghoff, H.~Yang, J.~Ebbers, F.~G. Germain, G.~Wichern, and J.~Le~Roux, ``Keeping the balance: Anomaly score calculation for domain generalization,'' in \emph{Proc. IEEE International Conference on Acoustics, Speech and Signal Processing (ICASSP)}, 2025, pp. 1--5.

\bibitem{wilkinghoff2026clusterexit}
K.~Wilkinghoff, G.~Wichern, J.~Le~Roux, and Z.-H. Tan, ``Mind the gap: Detecting cluster exits for robust local density-based score normalization in anomalous sound detection,'' \emph{arXiv preprint arXiv:2602.18777}, 2026.

\bibitem{vovk2005algorithmic}
V.~Vovk, A.~Gammerman, and G.~Shafer, \emph{Algorithmic Learning in a Random World}.\hskip 1em plus 0.5em minus 0.4em\relax Springer, 2005.

\bibitem{laxhammar2015}
R.~Laxhammar and G.~Falkman, ``Inductive conformal anomaly detection for sequential detection of anomalous sub-trajectories,'' \emph{Annals of Mathematics and Artificial Intelligence}, vol.~74, pp. 67--94, 2015.

\bibitem{angelopoulos2021conformal}
A.~N. Angelopoulos and S.~Bates, ``Conformal prediction: A gentle introduction,'' \emph{Foundations and Trends in Machine Learning}, vol.~16, no.~4, pp. 494--591, 2023.

\bibitem{ding2023classconditional}
T.~Ding, A.~N. Angelopoulos, S.~Bates, M.~I. Jordan, and R.~J. Tibshirani, ``Class-conditional conformal prediction with many classes,'' in \emph{Advances in Neural Information Processing Systems (NeurIPS)}, 2023.

\bibitem{gibbs2023conditional}
I.~Gibbs, J.~J. Cherian, and E.~J. Cand{\`e}s, ``Conformal prediction with conditional guarantees,'' \emph{Journal of the Royal Statistical Society Series B: Statistical Methodology}, vol.~87, no.~4, pp. 1100--1126, 2025.

\bibitem{ibm2026adaptiveconformal}
N.~Martinez~Gil, F.~O'Donncha, W.~M. Gifford, N.~Zhou, D.~C. Patel, and R.~Vaculin, ``Adaptive conformal anomaly detection with time series foundation models for signal monitoring,'' in \emph{Proc. International Conference on Learning Representations (ICLR)}, 2026.

\bibitem{weightedconformal2026}
O.~Hennh{\"o}fer and C.~Preisach, ``Between resolution collapse and variance inflation: Weighted conformal anomaly detection in low-data regimes,'' 2026.

\bibitem{ress2026conformal}
S.~Yuan, J.~Li, C.~Wang, and X.~Zhang, ``Conformal machine learning for reliable anomaly detection in industrial cyber-physical systems,'' \emph{Reliability Engineering \& System Safety}, vol. 274, p. 112417, 2026.

\bibitem{setvalued2025}
R.~Tsibulsky, D.~Nevo, and U.~Shalit, ``Set valued predictions for robust domain generalization,'' in \emph{Proc. International Conference on Machine Learning (ICML)}, 2025.

\bibitem{guan2023localized}
L.~Guan, ``Localized conformal prediction: a generalized inference framework for conformal prediction,'' \emph{Biometrika}, vol. 110, no.~1, pp. 33--50, 2023.

\bibitem{hore2023localweights}
R.~Hore and R.~F. Barber, ``Conformal prediction with local weights: randomization enables robust guarantees,'' \emph{Journal of the Royal Statistical Society Series B: Statistical Methodology}, vol.~87, no.~2, pp. 549--578, 2025.

\bibitem{cestnik1990}
B.~Cestnik, ``Estimating probabilities: {A} crucial task in machine learning,'' in \emph{Proc. 9th European Conference on Artificial Intelligence (ECAI)}, 1990, pp. 147--149.

\bibitem{gelman2007arm}
A.~Gelman and J.~Hill, \emph{Data Analysis Using Regression and Multilevel/Hierarchical Models}.\hskip 1em plus 0.5em minus 0.4em\relax Cambridge University Press, 2007.

\bibitem{metaod2021}
Y.~Zhao, R.~A. Rossi, and L.~Akoglu, ``Automatic unsupervised outlier model selection,'' in \emph{Advances in Neural Information Processing Systems (NeurIPS)}, 2021.

\bibitem{goswami2023uoms}
M.~Goswami, C.~Challu, L.~Callot, L.~Minorics, and A.~Kan, ``Unsupervised model selection for time-series anomaly detection,'' in \emph{Proc. International Conference on Learning Representations (ICLR)}, 2023.

\bibitem{sugiyama2007iwcv}
M.~Sugiyama, M.~Krauledat, and K.-R. M{\"u}ller, ``Covariate shift adaptation by importance weighted cross validation,'' \emph{Journal of Machine Learning Research}, vol.~8, no.~35, pp. 985--1005, 2007.

\bibitem{you2019dev}
K.~You, X.~Wang, M.~Long, and M.~Jordan, ``Towards accurate model selection in deep unsupervised domain adaptation,'' in \emph{Proc. International Conference on Machine Learning (ICML)}, ser. Proceedings of Machine Learning Research, vol.~97, 2019, pp. 7124--7133.

\bibitem{saito2021snd}
K.~Saito, D.~Kim, P.~Teterwak, S.~Sclaroff, T.~Darrell, and K.~Saenko, ``Tune it the right way: Unsupervised validation of domain adaptation via soft neighborhood density,'' in \emph{Proc. IEEE/CVF International Conference on Computer Vision (ICCV)}, 2021, pp. 9164--9173.

\bibitem{harada2021toyadmos2}
N.~Harada, D.~Niizumi, D.~Takeuchi, Y.~Ohishi, M.~Yasuda, and S.~Saito, ``{ToyADMOS2}: Another dataset of miniature-machine operating sounds for anomalous sound detection under domain shift conditions,'' in \emph{Proc. Detection and Classification of Acoustic Scenes and Events (DCASE) Workshop}, 2021.

\bibitem{dohi2022mimiidg}
K.~Dohi, T.~Nishida, H.~Purohit, R.~Tanabe, T.~Endo, M.~Yamamoto, Y.~Nikaido, and Y.~Kawaguchi, ``{MIMII DG}: Sound dataset for malfunctioning industrial machine investigation and inspection for domain generalization task,'' in \emph{Proc. Detection and Classification of Acoustic Scenes and Events (DCASE) Workshop}, 2022.

\bibitem{dcase2026task2}
T.~Nishida, N.~Harada, D.~Takeuchi, D.~Niizumi, K.~Imoto, K.~Dohi, H.~Purohit, T.~Endo, and Y.~Kawaguchi, ``Description and discussion on {DCASE} 2026 challenge task 2: Noise-aware unsupervised anomalous sound detection for machine condition monitoring,'' 2026.

\end{thebibliography}

\end{document}